# Influence of the growth interface shape on the defect characteristics in the facet region of 4H-SiC single crystals

Matthias Arzig [1*], Michael Salamon [2], Ta Ching Hsiao[3], Norman Uhlmann[2] and Peter J. Wellmann[1]

1) Crystal Growth Lab, Materials Department 6 (i-meet), University of Erlangen-Nürnberg (FAU), 91058 Erlangen, Germany (matthias.arzig@fau.de; peter.wellmann@fau.de)

2) Fraunhofer Institute for Integrated Circuits, Development Center for X-Ray Technology (EZRT), 90768 Fürth, Germany (michael.salamon@iis.fraunhofer.de; norman.uhlmann@iis.fraunhofer.de)

3) Industrial Technology Research Institute of Taiwan (ITRI), 195, Sec 4, Chung Hsing Rd., Chutung, Hsinchu 31040, Taiwan (TaChingHsiao@itri.org.tw)

* corresponding author: matthias.arzig@fau.de

Abstract

Two 75mm 4H-SiC single crystals are grown by the physical vapor transport (PVT) technique, using different insulation materials. The insulation material of higher thermal conductivity led to an increased radial temperature gradient. The evolution of the growth front was monitored using the in-situ computed tomography (CT). A slightly bent growth interface and a bigger facet are formed during the growth applying a lower radial temperature gradient while a smaller facet and steeper crystal flanks are formed in the case of the larger radial temperature gradient. Micropipes are deflected laterally by large surface steps on the steep crystal flanks and a reduction of threading edge dislocations by 60% is revealed by KOH defect etching.

1. Introduction

The growth via the physical vapor transport (PVT) method has matured to become the standard technique for the industrial production of silicon carbide (SiC) bulk material [1-4]. Continuous improvements of the wafer quality make micropipe (MP) free material commercially available and also overall defect densities are reduced. While in the PVT growth the process conditions are adjusted in order to reduce the generation of new defects as far as possible, with the solution growth method a distinct reduction of defects is possible [5]. It is believed that due to high surface steps threading defects are deflected into the basal plane. With our study we want to show that in the PVT growth of SiC similar conditions for the reduction of defects can be achieved, too. Using our in-situ computed tomography (CT) system we can observe the effect of changed process conditions directly during the growth. The influence of differently shaped growth interfaces on the surface morphology is shown and mechanisms for defect reduction are discussed in detail.

## 2. Experimental

Two 75 mm 4H-SiC crystals A and B were grown by the PVT method using an argon background pressure of 20mbar. In order to stabilize the 4H-Polytype, nitrogen is added into the gas phase. The carbon crucible is heated up inductively by a water cooled coil placed around the reaction chamber. Crystal A was grown using a powder source material produced at FAU, exhibiting a porosity of 0.66 while crystal B was grown with a powder source material provided by ITRI, featuring a similar porosity of 0.65. The growth is carried out on the C face of the 4H-SiC seed mounted at the top inside the crucible. For the growth of crystal B an insulation material with a higher thermal conductivity is used than for the growth of crystal A. As the thermal conductivity of the insulation material for crystal B is around 3 times higher, more power is necessary in order to reach the required temperatures for the growth process. The measured temperatures on top of the crucible were 2140°C for crystal A and 2095°C for crystal B.

The growth process was monitored using our in-situ CT system [6]. The growth setup is imaged with X-rays (125kV tungsten anode tube) on a digital flat panel detector (PaxScan 2520 D/CL, Varian Medical Systems). 200 X-Ray images are taken from different angles enabled by the 360° rotation of the growth setup over a time period of around 40 min. From this dataset of 2D-Projections, the 3D volume is calculated with the Feldkamp-Algorithm. CT measurements are conducted several times during the experiment and enabling a detailed look on how the mass distribution inside the crucible changes and how the crystal evolves during growth.

From the resulting crystals four wafers are cut from the curved crystal growth front close to the facet. After polishing, the wafers are etched in molten KOH for 5 minutes at a temperature of 510°C as described in [7].

## 3. Results

In Figure 1 the evolution of the growth experiments is depicted. The sublimation of the source material runs similar for both experiments leading to a growth rate of 340 µm/h for crystal A and 380 µm/h for crystal B. The shape of the growth front evolves differently in the two experiments. crystal A grows with a slightly convex growth interface. The growth front of crystal B is bent much stronger during the growth process, leading to steeper crystal flanks and a smaller facet. The difference in the developed crystal shape is due to the different insulation materials used for the experiments. In order to achieve the required temperatures much more heating power is needed for experiment B. Therefore, the crucible sidewalls are hotter leading to higher radial thermal gradients at the seed surface. This was validated by numerical simulation. In order to calculate the temperature field at the end of the growth phase, the mass distribution inside the crucible obtained from the in-situ CT data is taken into account for the numerical model as described in [8] . As depicted in Figure 2 the radial thermal gradient in the facet area is less pronounced for crystal A compared to crystal B. The larger radial thermal gradient in experiment B leads to the development of a smaller facet and steeper crystal flanks. Due to the larger inclination of the crystal flanks with respect to the facet plane the surface morphology changed, too. Figure 3 shows the surface morphology of the two crystals found at the crystal flanks next to the facet. While the surface of crystal A seems rather smooth, distinct macroscopic steps are formed on the surface of crystal B.

In order to evaluate the influence of the different growth morphologies on the defect distribution, KOH defect etching was performed. An area of 2x2 mm$^2$ adjacent to the facet is evaluated. It is ensured that the corresponding areas lying on top of each other in the crystal are analyzed in the wafers. This allows evaluating how the different step flow morphologies interact with the defects during the growth process. In Table 1 the results of the defect analysis are listed. For crystal A the number of threading edge dislocations (TED) is slightly increased by 7% , while the number of threading screw dislocations (TSD) and the basal plane dislocations (BPD) are reduced by 6% and 13%, respectively. For crystal B the number of BPDs increased by 2% and the number of TSDs decreased by 13%. For the TED density a remarkable reduction of 61% is found, indicating a strong interaction of this defect with the big surface steps.

The large steps found on the surface also influenced the MPs in crystal B. In Figure 4 corresponding sections of the etched wafers lying on top of each other in the crystal are depicted. The darker areas at the top right of the images stem from pronounced doping in the area of the $[000\bar{1}]$ main facet. The included dotted red line approximates the edge of the facet.  The positions of the MPs in the undermost wafer (Figure 4a) are marked by red circles and this initial position is also included in the images of the wafers on top (b-d). Going from the bottom to the top it gets obvious that the position of the MPs is steadily shifted further away from the marked position. This shift is not depending on a crystallographic orientation but it seems the MPs are deflected laterally by the big surface steps in the direction of step flow, away from the main facet.

In contrast, the etch pits of MPs located in the main facet area remain on top of each other comparing the wafers in Figure 3 a to d. A detailed analysis of all MPs in the facet area reveals that a small fraction slightly changes its lateral position in the crystal. This position change is not a collective movement like it's the case for the MPs in the step flow area. In Figure 5 the position shift of two MPs in corresponding areas of the etched wafers lying on top of each other inside the main facet is shown. The etch pits of the surrounding MPs stay on their position in all images. The position changes of the MPs proceeds in distinct directions. In the series from Figure 5 a to d Laue measurements confirm that the lateral movement of the etch pits is restricted to the $\{1\bar{1}00\}$ prismatic planes in the $\langle 11\bar{2}0 \rangle$ directions.

## 4. Discussion

The large surface steps that formed during growth of crystal B visible in Figure 3 lead to the assumption that a mechanism similar to the dislocation conversion utilized in solution growth led to a defect reduction. Using the solution growth method the growth interface typically shows macrosteps formed due to step bunching [9]. The lateral movement of the macrosteps can convert TSDs into defects in the basal plane and therefore lead to an overall reduction of the defect densities [9]. The conversion of TSDs leads to the formation of stacking faults. The stacking faults propagate laterally and therefore leave the crystal [10]. It was also shown that the step flow of the macrosteps can convert TEDs into BPDs [5]. Sato et al showed that macrosteps with heights of 100-1800 nm also led to the transformation of TSDs into Frank type stacking faults during the PVT growth of SiC [11] . The formation of stacking faults by the deflection of dislocations in the PVT growth was also shown by Dudley et al. in [12]. It is conceivable that the macrosteps formed on the growth interface of crystal B led to the transformation of several threading dislocations into defects in the basal plane, leaving the crystal at the crystal flanks. The reduction of the TEDs by 60% is probably due to a conversion into BPDs. The large surface steps deflect some of the TEDs into BPDs, as it was also observed to occur in PVT grown SiC in [13]. In experiment B the large radial thermal gradient leads to a stable flow of the large steps from the facet edge to the rim of the crystal. The steep crystal flanks prevent the formation of growth spirals with a step flow in the opposite direction and therefore a back deflection of the BPD into a TED becomes unlikely. This leads to an overall reduction of the TEDs while the number of BPDs is not increased as they leave the crystal laterally.

The large surface steps also influenced the MPs propagating through the crystal. While in crystal A the micropipes propagate parallel to the c axis the large surface steps of crystal B deflect the MPs in the direction of the step flow. The bending of the MPs away from the facet in the direction of the step flow leads to the change of the position of their etch pits visible in Figure 4 a to d. The bending of micropipes in SiC grown by sublimation epitaxy is also reported by R. Yakimova et al in [14]. In the CVD growth of SiC on patterned substrates it was shown that high surface steps with a steep slope can lead to the bending of screw dislocations [15]. In the facet area of the crystal the surface morphology is characterized by smaller steps and growth spirals. Therefore, the MPs in this area are not deflected and remain parallel to the c-axis. Nevertheless it was shown that individual etch pits of MPs seem to be shifted along distinct crystallographic directions as visible in the wafer series depicted in Figure 5. It is noticeable that some MPs seem to repel MPs close to them. The MPs are strictly deflected along the $\{1\bar{1}00\}$ prismatic planes in the $\langle11\bar{2}0\rangle$ directions. The slip of a hollow core screw dislocation after growth is not very likely. More conceivable is an interaction of the MPs leading to an increased distance between them during the growth. As MPs with a burgers vector of equal sign repel each other [16] one of the MPs is bent away from the other in the $\{1\bar{1}00\}$ plane.

## 5. Conclusion

The influence of different growth morphologies on the defect distribution in 4H-SiC single crystal was investigated. Using insulation material with a higher thermal conductivity led to a more strongly bent crystal growth interface, as observed in the in-situ CT. Due to the higher radial temperature gradient a smaller facet and steeper crystal flanks evolved compared to the growth run using an insulation material of lower thermal conductivity. On the steep crystal flanks high surface steps were found to strongly interact with the crystal defects. Two different mechanisms for the bending of micropipes away from the $[000\bar{1}]$ direction were found. In the facet area the repulsive interaction of selected MPs led to a slight deflection restricted to the $\{1\bar{1}00\}$ prismatic plane in the $\langle11\bar{2}0\rangle$ direction. In the step flow area the MPs are deflected collectively in the direction of the step flow by big surface steps. The large surface steps are also believed to convert a large amount of TEDs into BPDs leading to a reduction of this defect during the growth process. For the growth of large SiC boules the application of large radial gradients for defect reduction seems not appropriate as large radial strain will lead to the generation of new defects. Therefore, the application of an off-axis angle to produce a surface morphology featuring large steps would be a better method. Measurements of the angle between the facet plane and the crystal flanks, as depicted in Figure 6, reveal an inclination angle of 10° directly around the rim of the facet and going up to 20° further away at the steepest regions. The growth on seeds cut with a large off-axis angle between 10-20° can be a promising approach to achieve defect reduction in PVT grown SiC by large surface steps, while avoiding large radial thermal gradients.

Acknowledgements

The authors acknowledge funding by the DFG under the grant numbers WE2107/12-1 and UH246/4-1

Figures

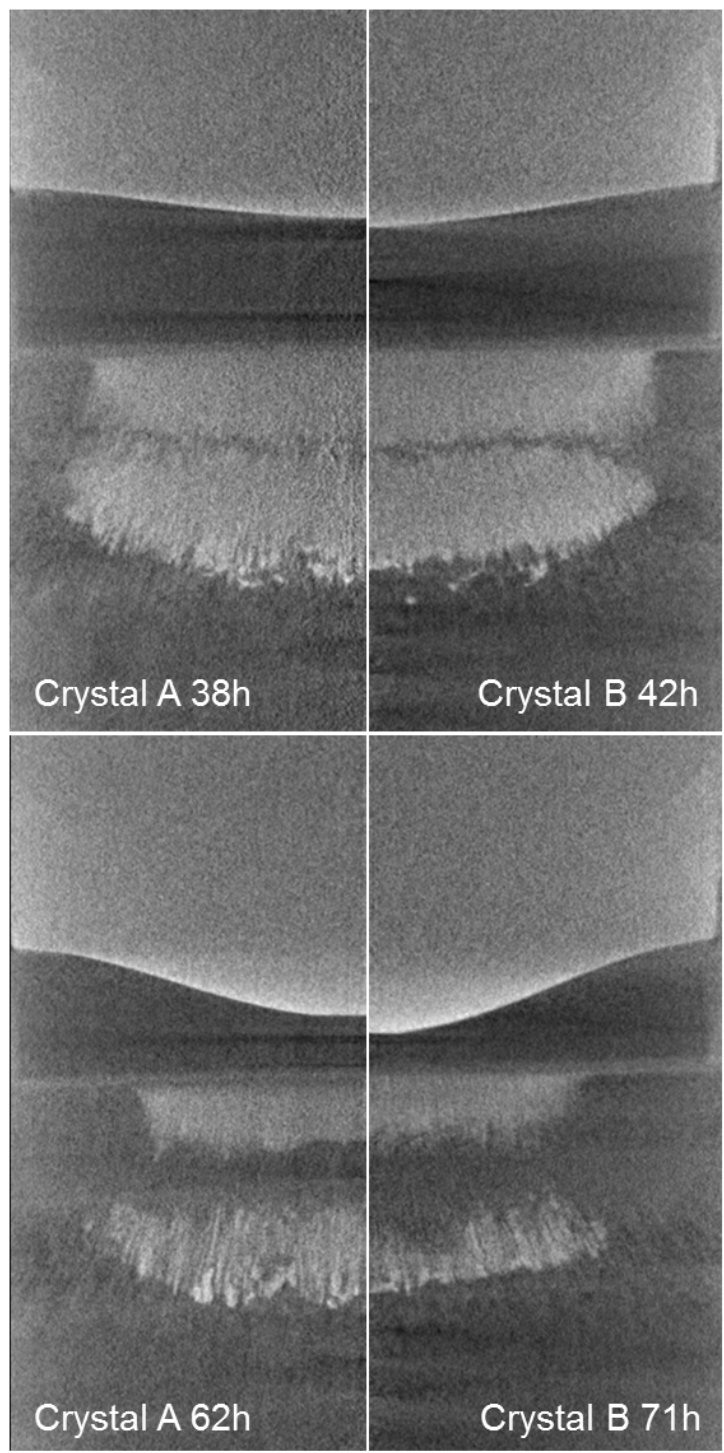

Figure 1: In-situ CT Images taken during the growth Experiments. The left side shows the evolution of the growth front of Crystal A at growth times of 38h at the top and 62h at the bottom. On the right side the evolution of the growth front of crystal B is depicted at growth times of 42h at the top and 71h at the bottom.

Single column fitting image

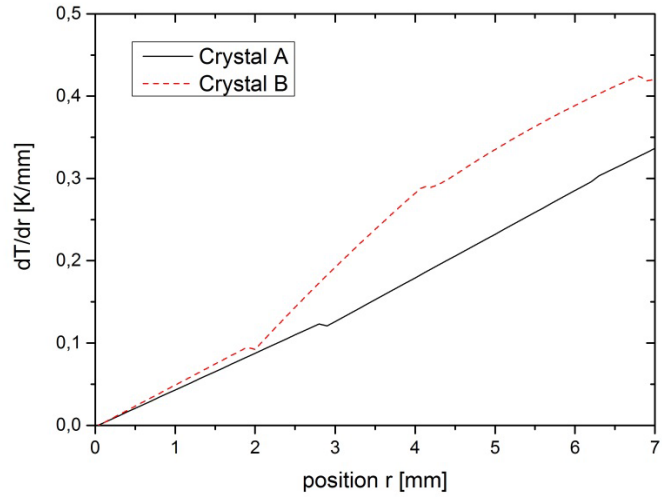

Figure 2: Radial thermal gradients close the facet at the end of the growth experiments. The continuous curve illustrates the thermal gradient dependent on the radial position for crystal A. The dotted line denotes the radial thermal gradient of crystal B.

Single column fitting image

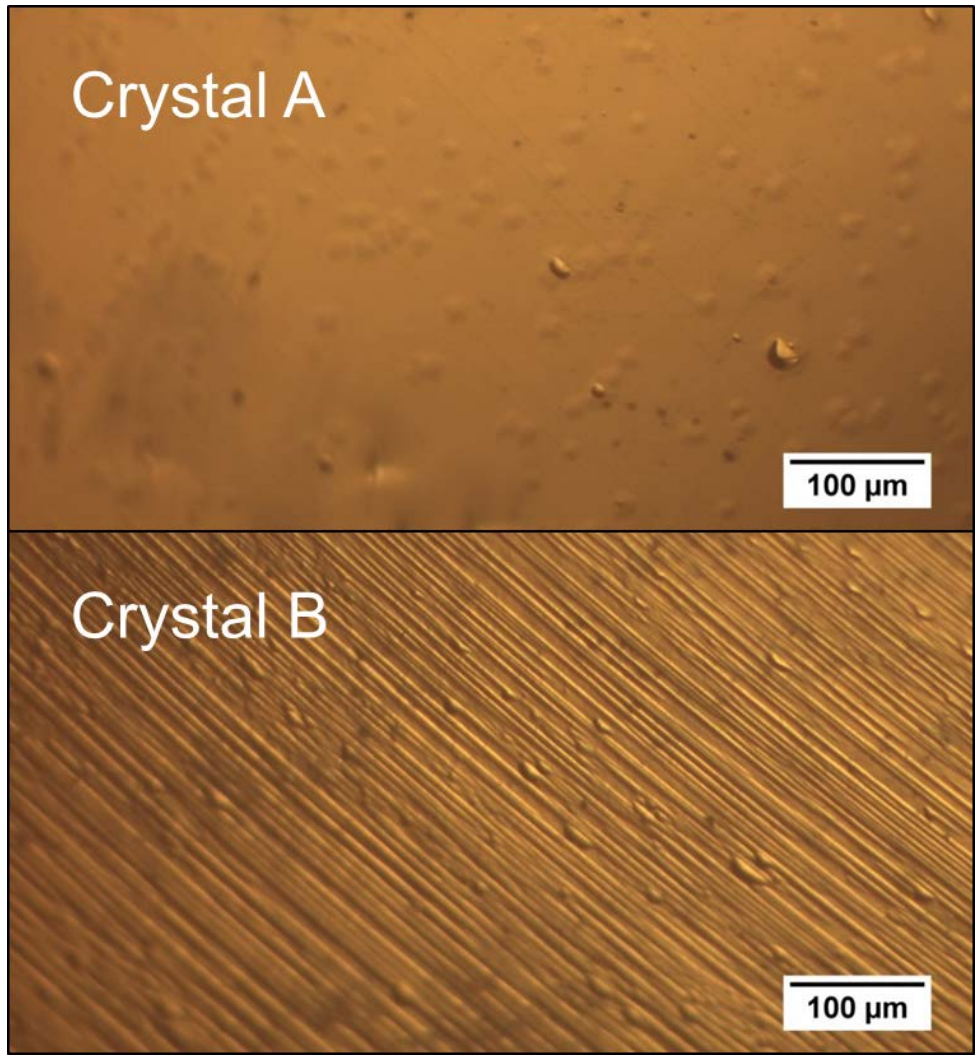

Figure 3 Optical Microscopy images taken from the crystal surfaces in the step flow growth area at the crystal flanks close to the facet. The top shows the surface morphology for crystal A and the bottom shows the surface morphology of crystal B.

Single column fitting image

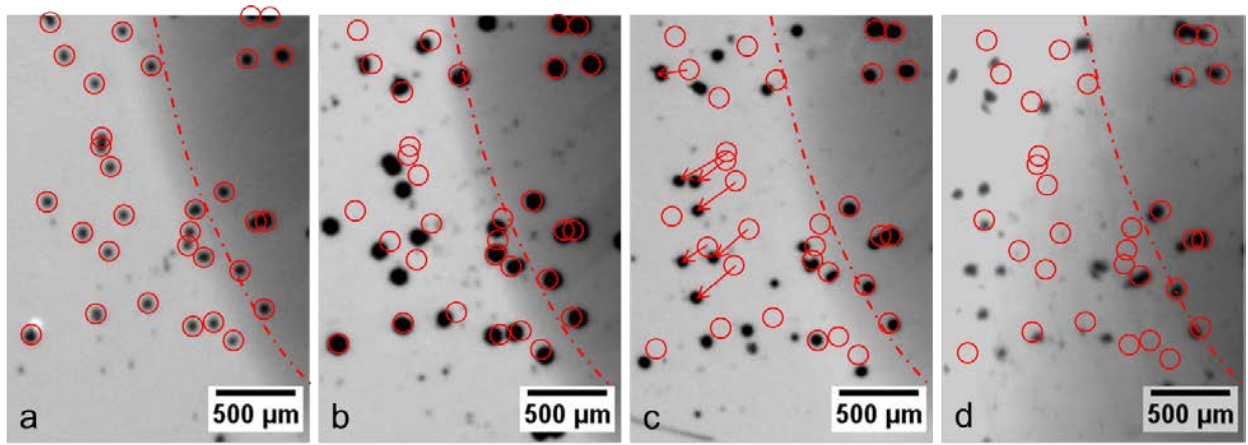

Figure 4 transmitted light images of the etched wafers from crystal B. The images are taken from corresponding areas lying on top of each other in the crystal. The letters a-d denote the position of the wafer in the crystal. Where the letter a denotes the undermost wafer and d the topmost wafer of the stack. The darker area represents the position of the main facet during growth due to pronounced Nitrogen incorporation. The red circles mark the position of the micropipes in Wafer number 4 in order to illustrate the shift of their position.

Double column fitting image

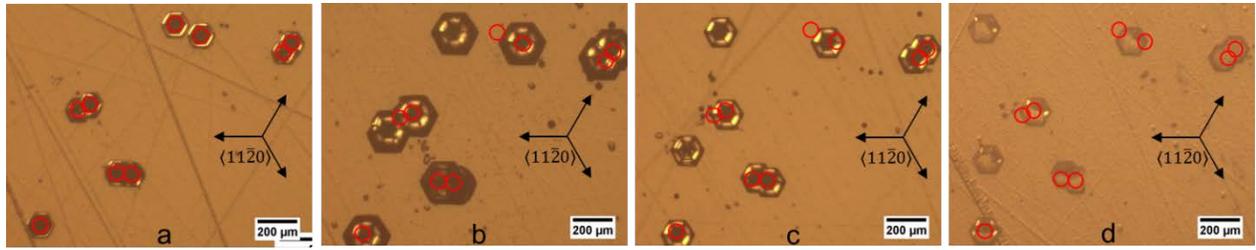

Figure 5 Nomarsky optical microscopy image of the etched wafers at corresponding areas inside the facet region of crystal B. The images from left to right correspond to the wafer position in the crystal from the bottom to the top.

Double column fitting image

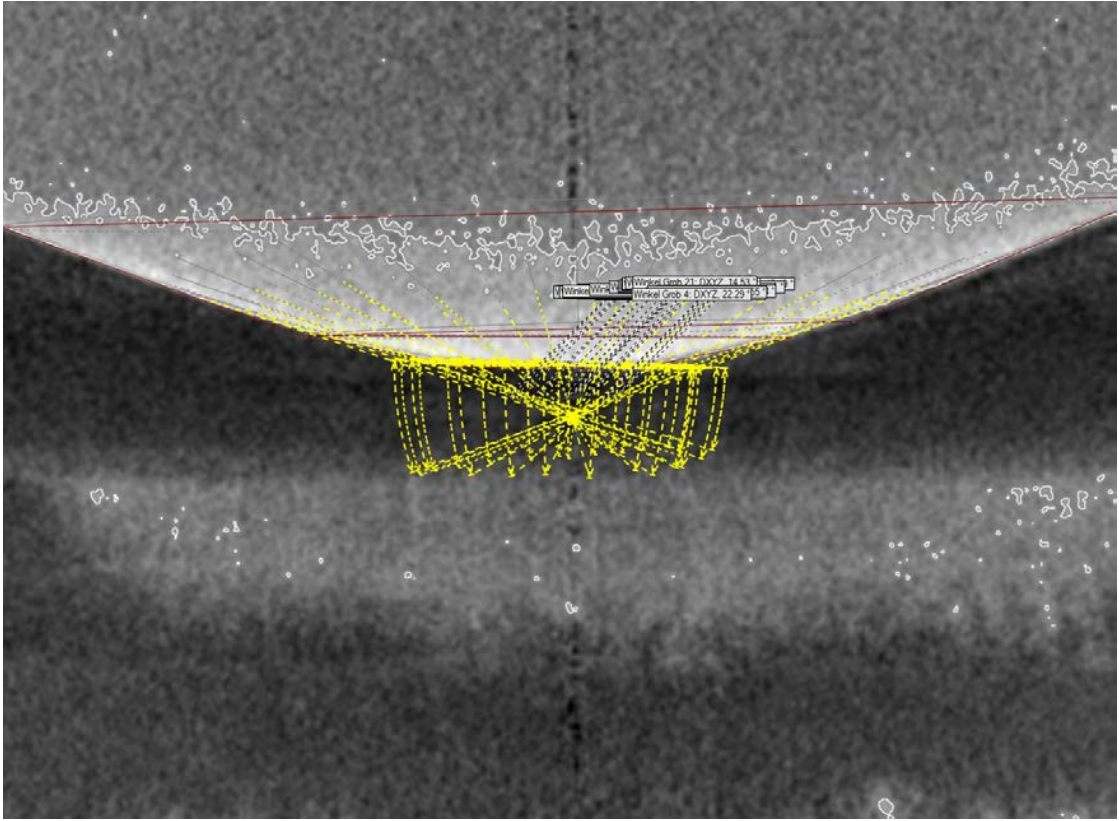

Figure 6 Measurement of the inclination angle between the facet plane and the crystal flanks in the in-situ CT Data.

Single column fitting image

Tables

Table 1 Etch pit densities for Crystal A and B counted for two Wafer portions lying on top of each other in the crystal volumes Wafer #1 denotes the wafer closer to the crystal growth front while Wafer #2 denotes the wafer closer to the seed.

|  | TED [1/cm$^2$] | TSD [1/cm$^2$] | BPD [1/cm$^2$] |
| --- | --- | --- | --- |
| Crystal A Wafer #1 | $(2.33 \pm 0.03)$E4 | $(0.94 \pm 0.02)$E3 | $(4.58 \pm 0.05)$ E3 |
| Crystal A Wafer #2 | $(2.18 \pm 0.02)$E4 | $(1.00 \pm 0.03)$E3 | $(5.27 \pm 0.05)$ E3 |
| Crystal B Wafer #1 | $(0.44 \pm 0.08)$E4 | $(0.88 \pm 0.29)$E3 | $(3.14 \pm 0.61)$ E3 |
| Crystal B Wafer #2 | $(1.12 \pm 0.30)$E4 | $(1.00 \pm 0.33)$ E3 | $(3.07 \pm 0.58)$ E3 |